\documentstyle[aps,preprint,prc,epsfig,12pt]{revtex}
\newcommand{\be}{\begin{equation}}
\newcommand{\ee}{\end{equation}}
\newcommand{\bd}{\begin{displaymath}}
\newcommand{\ed}{\end{displaymath}}
\newcommand{\ba}{\begin{eqnarray}}
\newcommand{\ea}{\end{eqnarray}}
\newcommand{\ket}[1]{| {#1} \rangle}
\newcommand{\bra}[1]{\langle {#1} |}
\newcommand{\ave}[1]{\langle {#1} \rangle}
\tightenlines

\begin{document}

\title{Symmetry conserving non-perturbative s-wave renormalization
of the pion in hot and baryon dense medium} 

\author{Z. Aouissat$^1$ and M. Belkacem$^2$}

\address{
$^1$ Institut f\"{u}r Kernphysik, Technische Hochschule 
Darmstadt, Schlo{\ss}gartenstra{\ss}e 9, \\
D-64289 Darmstadt, Germany. \\
$^2$ Institut f\"{u}r Theoretische Physik, J. W. Goethe-Universit\"{a}t, \\
D-60054 Frankfurt am Main, Germany
}

\date{ \today }
\maketitle

\begin{abstract}
A non-perturbative s-wave renormalization of the pion in a hot and baryon
rich medium is presented.  This approach proceeds via a mapping of the 
canonical pion into the axial Noether's charge. The mapping was made dynamical
in the Hartree-Fock-Bogoliubov random phase  approximation (HFB-RPA).
It is shown that this approach, while order mixing,  is still symmetry 
conserving both in the baryon free and baryon rich sectors, at zero as well
as  finite temperature. The systematic character of this approach is 
emphasized and it is particularly argued that it may constitute an interesting 
alternative for the non-perturbative assessment of the nuclear matter 
saturation properties.

\end{abstract}

{
\vskip 2\baselineskip
 
}

\newpage


\section{Introduction}

The understanding of the nuclear equation of state for hot and dense matter 
is one of the fiercely followed objective in the nuclear problem. 
Early studies although  successful  in
quantitatively assessing the nuclear matter ground-state properties, 
rely heavily on phenomenological input, as the nuclear two body interaction
for instance \cite{pand79,frie81}. These 
bear very little to the nowadays admitted fundamental theory of the strong
interaction, the quantum chromodynamics (QCD). 
Due to its strong coupling limit in the low energy regimes, QCD is in fact
intractable as a theory of its fundamental degrees of freedom. 
On the other hand, QCD has also symmetries which are as much fundamental. 
Their realization through effective hadronic theories seems to be a wise
compromise and a strategy adopted in the so-called quantum hadrodynamics (QHD)     
(see \cite{sero97} for a review). 

Of particular interest is chiral symmetry. Its nonlinear realization, 
through chiral perturbation theory  (ChPT)  for instance,
has proven  successful all over the past decade in understanding the 
low energy pion physics. This, however, falls short of answering 
some important questions as how to deal with highly collective structures
like bound states or resonances and how to fulfill
the fundamental unitarity condition.      
Thus one realizes that besides QCD's symmetries, 
the QHD inherits as well the difficult
aspects of the strongly interacting fields. Any reasonable handling
of these theories requires therefore an adequate treatment of the 
complicated vacuum structure. This is addressed in general through 
non-perturbative methods. The symmetry, however,  
imposes stringent constraints on the way the evaluation 
of the dynamics is conducted. Therefore one is called to be very selective
in choosing non-perturbative approaches that help in gathering the
dynamics without destroying the symmetry. 

This task, although trivial
in the case of the coupling constant perturbation (CCP), 
is full of subtleties and requires in fact  much more effort from within the 
strong interaction physics community. In this regard, as an educated 
non-perturbative approach, the concept of symmetry conserving 
dynamical mapping (SCDM) was introduced in \cite{aoui98}.
The idea is to map the original Fock-space, created at the quantization 
and which supports the CCP approach, to some ideal Fock-space selected
via a given symmetry conserving mapping (SCM). The next step is to make a 
projection onto the physical Fock-space which is a subspace
of the ideal Fock-space. The so-called projection is realized by making 
the SCM dynamical\footnote{Further precautions need, of course, 
to be observed in projecting out unphysical states. The necessary
technics to address this question are well available in the 
literature \cite{ring80}.}. 

There are various 
advantages in performing non-perturbative calculations
following this scheme.  The most interesting aspect is certainly the 
possibility of tracking all kind of transformations taking place in 
the  Fock-space while gathering the dynamics \cite{aoui98}. 
In particular the
asymptotic particles, which are highly dressed by the dynamics,  
have also well defined normalizable states in a well defined Fock-space.       
This is a clear departure from some traditional non-perturbative approaches
in which one loses this precious insight.
Probably the difficult task in this picture is to actually find the
right symmetry conserving mapping. 
 
Beside the $1/N$ expansion approach which was sorted-out
using Holstein-Primakoff mapping, an other SCM, based on a 
field-to-current mapping, was presented in \cite{aoui98,aoui96}. 
That this second mapping is systematic
in handling complicated situations was  proven in \cite{aoui98-a} within the
chirally invariant $SU(3)\times SU(3)$ Gellmann-Levy lagrangian.  
It will be further
consolidated here in considering a baryon rich environment at finite
temperature within the $SU(2)\times SU(2)$ version of the model. 
It will be also made clear later on that the field-to-current mapping is   
a robust SCDM which is able to handle any complicated
dynamics as long as the symmetry is linearly 
realized in the lagrangian.  
In what follows the field-current mapping will be made dynamical 
in the Hartree-Fock-Bogoliubov-RPA scheme. This was the  approximation
adopted previously in \cite{aoui96,aoui98-a}. 
Higher approximations are also possible
and these will soon be reported upon \cite{aoui99}.   

This paper is organized  as follows: We briefly review in section III
the effective  chiral model used all over the paper and
spell out the basic rule of the game, namely the preservation of the
lowest chiral ward identity throughout our non-perturbative treatment. 
This will be realized via an HFB-RPA scheme. Therefore, we first introduce    
in  subsection-A a Bogoliubov rotation and perform a self-consistent mean 
field calculation
which prepares the HFB basis. The RPA fluctuations build on the HFB basis
will be evaluated in subsection-B. It will be shown that the Goldstone
theorem is fulfilled exactly. In section IV, the finite temperature
and finite chemical potential extensions of this approach are made. 
Finally, the conclusions are drawn in section V.

\section{THE MODEL}

As a starting point, we recall the lagrangian density of the $SU(2)\times SU(2)$ 
linear $\sigma$-model with nucleonic degrees of freedom. 
To realize an explicit chiral symmetry breaking in the PCAC sense,
one adds the usual linear term in the sigma field.
\be
\hspace*{-0.3cm}{\cal L}_{\sigma} =  \bar{\psi} \left[ i\not\partial 
        + g  \left( \hat{\sigma} + i \vec{\tau} \vec{\pi} 
        \gamma_5 \right) \right] \psi
        + \frac{1}{2}\left[(\partial_{\mu}\vec{\pi})^2 + 
        (\partial_{\mu}\hat{\sigma})^2 \right] - \frac{\mu^2}{2}
        \left[\vec{\pi}^2 +\hat{\sigma}^2 \right] -\frac{\lambda^2}{4}
        \left[\vec{\pi}^2 +\hat{\sigma}^2 \right]^2 + c \hat{\sigma}
\label{eq1}
\ee
Here $\psi$, $\hat{\sigma}$ and $\pi$ represent the bar nucleon, sigma,
and pion fields, respectively. To fix our notations for latter use, 
these are given below, in terms of their respective creation and
annihilation operators, by the usual plane wave expansions
\ba
\psi({\bf x}) &=& \int \frac{d{\vec p}}{\sqrt{(2\pi)^3 2E_p}}
\sum_{r}
\left[ u({\bf p},r) e^{-i{\bf px}} c_{{\bf p}r} + 
v({\bf p},r) e^{i{\bf px}} d^+_{{\bf p}r} \right] , 
\nonumber \\
\pi_i({\bf x}) &=& \int \frac{d{\vec p}}{\sqrt{(2\pi)^3 2\omega_p}}
\left( e^{i{\bf px}} a_{{\bf p}i} + 
e^{-i{\bf px}} a^+_{{\bf p}i} \right) , 
\quad
\hat{\sigma}({\bf x}) = \int \frac{d{\vec p}}{\sqrt{(2\pi)^3 2\omega_p}}
\left( e^{i{\bf px}} b_{{\bf p}} + 
e^{-i{\bf px}} b^+_{{\bf p}} \right)
\label{eq2}
\ea
The on-shell energies are taken to be  $\omega_p  = \sqrt{\mu^2 + p^2}$ for 
the bosons, and $E_p =p$ for the massless fermions. The definition of 
the bi-spinors $u({\bf p},r)$ and $v({\bf p},r)$ considered here is the one which
takes care of this limiting case \cite{itzy80}. 
  
As stated in the introduction, the  aim here is to conduct
a non-perturbative treatment of the model that goes 
beyond the usual coupling constant perturbation (CCP)
as well as the semi-classical 
approaches. These two  are,  presently, the most  
extensively used approaches in the study of the dynamics of these QCD inspired 
hadronic models.
Before embarking in this, 
we recall the commonly used classical mean-field approximation. In this case,
the masses of the different particles are given by
\be
m_{\pi}^2 = \mu^2 + \lambda^2\langle \hat{\sigma} \rangle^2~,~~
m_{\sigma}^2 =  \mu^2 + 3\lambda^2\langle \hat{\sigma} \rangle^2~,~~
m_F^2 = - g\langle \hat{\sigma} \rangle~,~~
c = \mu^2\langle \hat{\sigma} \rangle + 
\lambda^2\langle \hat{\sigma} \rangle^3
\label{eq3}
\ee
where $\langle \hat{\sigma} \rangle$ denotes the vacuum expectation value of
the sigma field. At this level, the Goldstone nature of the pion, in the chiral limit
$(c =0)$, is manifest. Indeed, one can also verify that the Ward identity 
\be
-D_{\pi}^{-1}(0) = \frac{c}{\ave{\hat{\sigma}}}
\label{eq4}
\ee
is fulfilled. Of course, all higher Ward identities (WI) are also preserved 
by the CCP approach at each order of the perturbation,
and thus, at this lowest order too. 
 The present WI holds as well in the case of the semi-classical non-perturbative
approach \`a la $1/N$-expansion\footnote{N is the number of pion charges}.
In fact, in the large $N$-limit,
the condensate and the pion mass
are obtained as solutions of two coupled  BCS equations in the 
Hartree-Bogoliubov approximation (see for instance \cite{aoui98,aoui97}). 
The fluctuating part of the sigma field is found to be not involved at all in
building the variational ground state for the whole $1/N$-expansion approach. 
As such the sigma mass has an exclusively  perturbative character. 
Here too, the whole hierarchy of WI's, as in the case of the CCP,  is preserved.  
In the present paper, however, our ambition is rather modest. 
Our aim is simply to set up a non-perturbative approach, which transcends
the two approximations cited above, but  still preserves in a systematic way
the lowest WI in Eq.(\ref{eq4}), at all conditions of temperature and 
baryon density.

\subsection{HARTREE-FOCK-BOGOLIUBOV MEAN-FIELD}

The starting point for our mean field considerations is the following
Bogoliubov transformation for both nucleon and meson operators
\ba
\vec{\alpha}_{\bf p}^+ &=& u_{\pi}(p)\vec{a}_{\bf p}^+ - v_{\pi}(p) \vec{a}_{\bf
-p}~~;~~~
\beta_{{\bf p}}^+ = u_{\sigma}(p)b_{{\bf p}}^+ - v_{\sigma}(p) b_{{\bf -p}} - 
w_0 \delta({\bf p})
\nonumber \\
C_{{\bf p}r}^+ &=& u_F(p) c_{{\bf p}r}^+ - r v_F(p) d_{{\bf -p}r}~~;~~~
D_{{\bf p}r}^+ = u_F(p) d_{{\bf p}r}^+ + r v_F(p) c_{{\bf -p}r}
\label{eq5}
\ea
where $u_{\pi}(p)$, $u_{\sigma}(p)$, $v_{\pi}(p)$, $v_{\sigma}(p)$, $u_F(p)$
and $v_F(p)$ are even
functions of their arguments. To account for a finite $\ave{\hat{\sigma}}$ 
condensate in the Goldstone phase, the Bogoliubov transformation is made 
inhomogeneous for the sigma field by introducing a c-number shift $w_0$.  
The canonicity of the above transformations is enforced by means of the 
following conditions : 
\be
u_{\pi}(p)^2 - v_{\pi}(p)^2 = 1~,\quad\quad
 u_{\sigma}(p)^2 - v_{\sigma}(p)^2 = 1~,\quad\quad
u_F(p)^2 + v_F(p)^2 = 1
\label{eq6}
\ee
Accordingly, the vacuum $\ket{0}$ of the theory at the classical mean field,
defined by: $ {\vec a}_{\pi}\ket{0} = b_{\sigma} \ket{0} = 
c\ket{0} = d \ket{0} = 0 $, 
is  unitarily transformed
to a new one $\ket{\Phi}$, such that
\bd
 {\vec \alpha}\ket{\Phi} = \beta \ket{\Phi} = 
C\ket{\Phi} = D \ket{\Phi} = 0~,
\ed
and  explicitly given in terms of the following squeezed vacuum
\be
\ket{\Phi}  =  
\exp \left[ \int d{\vec p} 
\left\{ z_{\pi}(p)\vec{a}_{\bf p}^+\vec{a}_{\bf -p}^+   + 
        z_{\sigma}(p)b_{{\bf p}}^+b_{{\bf -p}}^+     + 
        z_{F}(p) \sum_{r} r c_{{\bf p}r}^+ d_{{\bf -p}r}^+ \right\} + 
       \frac{w_0}{u_{\sigma}(0)} b_{{\bf 0}}^+ - \mathrm{h.c.} 
\right] \ket{0}~. 
\label{eq7}
\ee
where $z_{\pi}(p)=\mathrm{arcch}\mathnormal[u_{\pi}(p)]$, 
 $z_{\sigma}(p)=\mathrm{arcch}\mathnormal[u_{\sigma}(p)]$ and
$z_{F}(p) = \arccos[u_F(p)]$. 
The Hamiltonian of the model derived from Eq.(\ref{eq1}) 
and written in the quasi-particles basis reads:
\ba
{\cal H} &=& {\cal H}_0 \,+\,  \eta \left[ \beta_0 + \beta_0^+ \right] \,+\,
 \int d{\vec p}\,\, {\cal E}_F(p) \sum_{r}
\left[ C_{{\bf p}r}^+C_{{\bf p}r} + D_{{\bf p}r}^+D_{{\bf p}r} \right] 
\,+\, \int d{\vec p} \,\,{\cal E}_{\pi}(p) \vec{\alpha}_{\bf p}^+ 
\vec{\alpha}_{\bf p}
\nonumber \\
&+& \int d{\vec p} \,\,{\cal E}_{\sigma}(p)  \beta^+_{\bf p} \beta_{\bf p} 
\,+\,  \int d{\vec p}\, c_F(p) \sum_{r}\,r\,
\left[  C_{{\bf p}r}^+D_{{\bf -p}r}^+ + 
D_{{\bf -p}r}C_{{\bf p}r} \right] 
\nonumber \\
&+& \int d{\vec p}\,
     c_{\pi}(p) \left[ \vec{\alpha}_{\bf p}^+ \vec{\alpha}_{\bf -p}^+ +
   \vec{\alpha}_{\bf p} \vec{\alpha}_{\bf -p} \right]  \,+\,
 \int d{\vec p}\,  
   c_{\sigma}(p) \left[ \beta^+_{\bf p} \beta^+_{\bf -p} +
   \beta_{\bf p} \beta_{\bf -p} \right]  
\nonumber\\
&-& \int d{\bf x} : \left[ g \bar{\psi} \left( \sigma + 
i \vec{\tau} \vec{\pi} \gamma_5 \right) \psi \right] :    
\,+\,  \int d{\bf x} : \left[ \lambda^2 \ave{\hat{\sigma}} \sigma
  \left( \vec{\pi}^2 + {\sigma^2} \right) +
 \frac{\lambda^2}{4}\left( \vec{\pi}^2 + {\sigma^2} \right)^2
 \right] :
\label{eq8}
\ea
where $\sigma$ represents the shifted sigma-field: $\sigma =
\hat{\sigma} - \ave{\hat{\sigma}}$, and   $\ave{\hat{\sigma}}$ is 
the sigma condensate
which is given in term of the parameters of the Bogoliubov transformation
in Eq.(\ref{eq5}) by 
\bd
\ave{\hat{\sigma}}\,=\,\frac{\ave{b_0^+}+\ave{b_0}}{\sqrt{(2\pi)^3 2\mu}}
\, =\, \frac{(u_{\sigma}(0)+v_{\sigma}(0))(w_0\,+\,w_0^*)}{\sqrt{(2\pi)^3 
2\mu}}~.
\ed
The semicolons ":" in the expression (\ref{eq8}) of the hamiltonian  
denote operators normal ordering. Finally, the 
coefficients ${\cal H}_0$, $\eta$, 
${\cal E}_{F,\pi,\sigma}$ and $c_{F,\pi,\sigma}$ read explicitly
\ba
{\cal H}_0 &=& -\gamma \int \frac{d{\vec p}}{(2\pi)^3} \left[ E_p 
\left( u_F(p)^2 - v_F(p)^2 \right) - 
2g \ave{\hat{\sigma}} u_F(p)v_F(p) \right]  
\nonumber \\
&+&  \int \frac{d{\vec p}}{(2\pi)^3}\,\,
 \omega_p \left( 3 v_{\pi}(p)^2 +v_{\sigma}(p)^2+2 \right) 
\nonumber \\
&+& \frac{3\lambda^2}{4} \left[ I^2_{\sigma} +5 I^2_{\pi} 
+ 2I_{\sigma}I_{\pi} \right]
 + \frac{3 \lambda^2 \ave{\hat{\sigma}}^2}{2}\left[ I_{\sigma} + I_{\pi} \right]
 + \frac{\mu^2 \ave{\hat{\sigma}}^2}{2} + \frac{\lambda^2 \ave{\hat{\sigma}}^4}
 {4} - c\ave{\hat{\sigma}}~,
\label{eq10}
\ea
where $\gamma=4$ is the spin-isospin degeneracy of the nucleon, and

\ba
\eta &=& \sqrt{(2\pi)^3}\, \frac{u_{\sigma}(0) +v_{\sigma}(0)}{\sqrt{2 \mu}}
\left[ 3\lambda^2 \ave{\hat{\sigma}} I_{\sigma} + 3\lambda^2 \ave{\hat{\sigma}} I_{\pi} 
+ \lambda^2 \ave{\hat{\sigma}}^3 + \mu^2 \ave{\hat{\sigma}} -  c - g I_F\right]
~,
\nonumber\\
   c_F(p) &=& g\ave{\hat{\sigma}} \left( u_F(p)^2 - v_F(p)^2 \right) + 
   2E_p u_F(p)v_F(p) ~,
\nonumber \\
   c_{\pi}(p) &=&  \omega_p u_{\pi}(p)v_{\pi}(p) +
   \frac{\lambda^2}{2}
   \frac{ \left( u_{\pi}(p)+v_{\pi}(p) \right)^{2}}{2\omega_{p}}
    \left[ 5I_{\pi} + I_{\sigma} +  \ave{\hat{\sigma}}^2 \right] ~, 
 \nonumber\\
   c_{\sigma}(p) &=&  \omega_{p} u_{\sigma}(p)v_{\sigma}(p) +
 \frac{3 \lambda^2}{2}
   \frac{ \left( u_{\sigma}(p)+v_{\sigma}(p) \right)^{2}}{2\omega_{p}}
    \left[ I_{\pi} + I_{\sigma} +  \ave{\hat{\sigma}}^2 \right] ~,
\nonumber\\
 {\cal E}_F(p) &=& E_p \left( u_F(p)^2 - v_F(p)^2 \right) - 
 2g\ave{\hat{\sigma}}u_F(p)v_F(p) ~,
\nonumber \\
 {\cal E}_{\pi}(p) &=&  \omega_{p} \left( u_{\pi}(p)^2 + v_{\pi}(p)^2 \right) +
   \lambda^2
   \frac{ \left( u_{\pi}(p)+v_{\pi}(p) \right)^{2}}{2\omega_{p}}
    \left[ 5I_{\pi} + I_{\sigma} +  \ave{\hat{\sigma}}^2 \right] ~,
 \nonumber\\
 {\cal E}_{\sigma}(p) &=&  \omega_{p} \left( u_{\sigma}(p)^2 + 
 v_{\sigma}(p)^2 \right) +
 3 \lambda^2
   \frac{ \left( u_{\sigma}(p)+v_{\sigma}(p) \right)^{2}}{2\omega_{p}}
    \left[ I_{\pi} + I_{\sigma} +  \ave{\hat{\sigma}}^2 \right] ~.
\label{eq11}
\ea
$I_{\pi}$, $I_{\sigma}$ and $I_{F}$ are expectation values on the squeezed vacuum
of bilinear forms of the pion, the sigma and the 
baryon fields, respectively. They read 
\ba
I_{\pi} &=& \int d{\bf x} \,\, \bra{\Phi} \pi_i({\bf x}) \pi_i({\bf x})\ket{\Phi}
\,=\,\int \frac{d{\vec p}}{(2\pi)^3}
   \frac{ \left( u_{\pi}(p)+v_{\pi}(p) \right)^2}{2\omega_p},
\nonumber\\
I_{\sigma} &=& \int d{\bf x} \,\, \bra{\Phi} \sigma({\bf x})\sigma({\bf x})\ket{\Phi}
\,=\,\int \frac{d{\vec p}}{(2\pi)^3}
   \frac{ \left( u_{\sigma}(p)+v_{\sigma}(p) \right)^2}{2\omega_p},
\nonumber\\
I_{F} &=& \int d{\bf x} \,\, \bra{\Phi} \bar\psi({\bf x}) \psi({\bf x}) \ket{\Phi}
= -\gamma \int  \frac{d{\vec p}}{(2\pi)^3} 2u_F(p)v_F(p)~. 
\label{eq12}
\ea
To fully fix the amplitudes $u_{\pi}$, $v_{\pi}$, $u_{\sigma}$, $v_{\sigma}$,
$u_F$, $v_F$, as well as the value of the
sigma condensate $\ave{\hat{\sigma}}$, we make use of the  Rayleigh-Ritz 
variational principle. 
Minimizing the ground state energy 
${\cal H}_0=\bra{\Phi} H \ket{\Phi}/\ave{\Phi | \Phi}$ while maintaining 
the canonicity condition of the Bogoliubov transformation 
Eq.(\ref{eq6}), one gets,  after a straightforward algebra,
the following set of equations

\begin{eqnarray} 
\left.\frac{\delta {\cal H}_0}{\delta u_{\pi}(p)}
\right|_{u_{\pi}^2 -v_{\pi}^2 =1} 
&=& \,\,0  \quad\quad \Rightarrow \quad\quad c_{\pi}(p)\,\,=\,\,0 ~,
\nonumber\\
\left.\frac{\delta {\cal H}_0}{\delta u_{\sigma}(p)}
\right|_{u_{\sigma}^2 -v_{\sigma}^2 =1}
&=& \,\,0  \quad\quad \Rightarrow \quad\quad c_{\sigma}(p)\,\,=\,\,0 ~,
\nonumber\\ 
\left.\frac{\delta {\cal H}_0}{\delta u_F(p)}\right|_{u_F^2 + v_F^2 =1}
&=& \,\,0  \quad\quad \Rightarrow \quad\quad c_{F}(p)\,\,=\,\,0 ~, 
\nonumber\\
\frac{\delta {\cal H}_0}{\delta \ave{\hat{\sigma}}} \quad\quad
&=& \,\,0  \quad\quad \Rightarrow \quad\quad  \eta \,\,=\,\,0 ~.             
\label{eq13}
\end{eqnarray}
From the  above, one sees that at the minimum of the HFB ground
state, the Hamiltonian in Eq.(\ref{eq8}) reduces to a sum of 
un-coupled quasi-particle modes (bilinear parts) and 
residual interactions  which are normal ordered with respect to the
squeezed vacuum. The set of equations in Eq.(\ref{eq13}) constitutes 
four coupled and self-consistent gap equations. 
Inserting the solutions of these into the definitions of the quasi-particle
energies, one gets the following:
 
\ba
{\cal E}_{\pi}(p) &=& \omega_p \left( u_{\pi}(p) - v_{\pi}(p) \right)^2 = 
\sqrt{\omega_p^2 +\lambda^2 
\left[ 5I_{\pi} +I_{\sigma} +\ave{\hat{\sigma}}^2 \right]}~,
\nonumber \\
{\cal E}_{\sigma}(p) &=& \omega_p \left( u_{\sigma}(p) - v_{\sigma}(p) \right)^2 = 
\sqrt{\omega_p^2 +3\lambda^2 
\left[ I_{\pi} +I_{\sigma} +\ave{\hat{\sigma}}^2 \right]}~,
\nonumber \\
{\cal E}_{F}(p) &=&  
\frac{\left( p^2 + g^2 \ave{\hat{\sigma}}^2 \right)}{E_p}
\left[u_F(p)^2 - v_F(p)^2\right]
=  \sqrt{p^2 + g^2 \ave{\hat{\sigma}}^2}~,
\nonumber \\ 
\frac{c}{\ave{\hat{\sigma}}} &=& \mu^2  - \frac{g}{\ave{\hat{\sigma}}} I_F + 3\lambda^2  (I_{\pi} + 
I_{\sigma}) + 
\lambda^2 \ave{\hat{\sigma}}^2~.  
\label{eq14}
\ea

\noindent
Finally, one deduces the BCS gap equations for the quasi-particle masses as well as 
the condensate 

\ba
{\cal E}_{\pi}^2(0) &=&\mu^2 
+ \lambda^2 \left[ 5I_{\pi} +I_{\sigma} + \ave{\hat{\sigma}}^2 \right]~,
\nonumber \\
{\cal E}_{\sigma}^2(0) &=& \mu^2 
+ 3\lambda^2 \left[ I_{\pi} +I_{\sigma} +\ave{\hat{\sigma}}^2 \right]~,
\nonumber \\
\ave{\hat{\sigma}} &=& - \frac{c}{-\mu^2} - \frac{g}{- \mu^2} I_F   
  + \frac{3\lambda^2 \ave{\hat{\sigma}}}{-\mu^2} I_{\pi} + \frac{3\lambda^2 
  \ave{\hat{\sigma}}}{-\mu^2} I_{\sigma} 
  + \frac{\lambda^2 \ave{\hat{\sigma}}^3}{-\mu^2} ~,
\nonumber \\
{\cal E}_F(0) &=& -g\ave{\hat{\sigma}}~.
\label{eq15}
\ea
Here $I_{\pi}$, $I_{\sigma}$ and $I_F$ are nothing but the 
one loop tadpoles of the fully dressed quasi-pion, quasi-sigma 
and quasi-fermion propagators, respectively.

\ba
I_{\pi}  &=&  \int \frac{d{\vec p}}{(2\pi)^3}
  \frac{1}{2\sqrt{p^2 + {\cal E}_{\pi}^2(0)}} ,
\nonumber \\ 
I_{\sigma} &=& \int \frac{d{\vec p}}{(2\pi)^3}
  \frac{1}{2\sqrt{p^2 + {\cal E}_{\sigma}^2(0)}} , 
\nonumber \\
I_F &=& -2 \gamma \int \frac{d{\vec p}}{(2\pi)^3} \frac{{\cal E}_F(0)}{2 \sqrt{p^2 
+ {\cal E}_F^2(0)}} 
\label{eq16}
\ea
The Feynmann diagrammatic representation of the  gap equations in
Eq.(\ref{eq15}) is displayed in Figure 1.
Through the  Yukawa coupling, the fermion is in fact minimally coupled 
to the rest of the gap equations. The gap equation for the fermion 
could be more tightly coupled if one considers, for instance, 
a chirally invariant quartic interaction among fermions
\`a la Nambu Jona-Lasinio. This will be briefly discussed at the end of the
paper.

At this point, a comment is in order. After inspection of the BCS gap equations 
Eq.(\ref{eq15}), it is clear that a solution with finite condensate
($\ave{\hat{\sigma}} \neq 0$) will lead to the following values for the masses 
of the quasi-pion and quasi-sigma :    
\ba
{\cal E}_{\pi}^2(0) &=& \frac{c + gI_F}{\ave{\hat{\sigma}}} + 
2\lambda^2 \left[ I_{\pi} -I_{\sigma} \right]
\nonumber \\
{\cal E}_{\sigma}^2(0) &=& \frac{c + gI_F}{\ave{\hat{\sigma}}} + 
2\lambda^2 \ave{\hat{\sigma}}^2
\label{eq17}
\ea
In this Nambu-Goldstone phase,  
these two modes are not degenerate as it should be. Therefore, the difference 
$I_{\pi} -I_{\sigma}$ as well as the fermionic tadpole $I_F$ are 
in fact finite. Thus in the chiral limit $(c \rightarrow 0)$, the Goldstone
mode seems to be absent from the spectrum of the theory, since ${\cal E}_{\pi}$
does not vanish. Consequently, one can check that the Ward identity
in Eq.(\ref{eq4}) is also violated. Since the condensate and 
the Goldstone mass are the building blocks for all
higher Ward identities, it is clear that 
the quasi-sigma mass is also unphysical in this approach. 
It appears then that the squeezed vacuum, apart
from the fact that it can accommodate a finite condensate, does neither have  
the right curvature for the sigma-like excitation, nor a valley 
along the pion-like excitation. In the following, we can only propose a  way
to resolve the second problem, namely correcting for the pion-like excitation
so as to fulfill the Goldstone theorem. This will have quantitative
consequences, as there will be important corrections to the ground
state energy.

\subsection{~~RPA  FLUCTUATIONS}

To proceed further, we recall the Hamiltonian in Eq.(\ref{eq8}). 
In the HFB basis, the latter takes the  form: 

\ba
{\cal H} &=& {\cal H}_0 +  
\int d{\vec p} \,{\cal E}_F(p) \sum_ {r}
\left[ C_{{\bf p}r}^+C_{{\bf p}r} +D_{{\bf p}r}^+D_{{\bf p}r} \right] 
+ \int d{\vec p}\, {\cal E}_{\pi}(p) \vec{\alpha}_{\bf p}^+ \vec{\alpha}_{\bf p}
     +
\int d{\vec p}\, {\cal E}_{\sigma}(p)  \beta^+_{\bf p} \beta_{\bf p}     
\nonumber \\
&-&   \int d{\bf x} : \left[ g \bar{\psi} \left( \sigma + 
i \vec{\tau} \vec{\pi} \gamma_5 \right) \psi \right] :
+  \int d{\bf x} : \left[ \lambda^2 \ave{\hat{\sigma}} \sigma
  \left( \vec{\pi}^2 + {\sigma^2} \right) +
 \frac{\lambda^2}{4}\left( \vec{\pi}^2 + {\sigma^2} \right)^2
 \right] :
\label{eq18}
\ea

Here the fields $\psi({\bf x})$, $\pi_i({\bf x})$ and $\sigma({\bf x})$  have 
now the following  plane wave expansions in the quasi-particle basis

\ba
\psi({\bf x}) &=& \int \frac{d{\vec p}}{\sqrt{(2\pi)^3 2{\cal E}_F(p)}}
\sum_{r}
\left[ u({\bf p},r) e^{-i{\bf px}} C_{{\bf p}r} + 
v({\bf p},r) e^{i{\bf px}} D^+_{{\bf p}r} \right] , 
\nonumber \\
\pi_i({\bf x}) &=& \int \frac{d{\vec p}}{\sqrt{(2\pi)^3 2{\cal E}_{\pi}(p)}}
\left( e^{i{\bf px}} \alpha_{{\bf p}i} + 
e^{-i{\bf px}} \alpha^+_{{\bf p}i} \right) , 
\nonumber \\
{\sigma}({\bf x}) &=& \int \frac{d{\vec p}}{\sqrt{(2\pi)^3 2{\cal E}_{\sigma}(p)}}
\left( e^{i{\bf px}} \beta_{{\bf p}} + 
e^{-i{\bf px}} \beta^+_{{\bf p}} \right)~,
\label{eq19}
\ea
where the harmonic modes are the respective quasi-particle energies.   

For a further gathering of the dynamics, we want to proceed by 
a perturbative diagonalization of the Hamiltonian\footnote{What is meant here 
is not the coupling constant perturbation. This word is  used here to highlight 
the additive character of this approach as opposed to a variational
self-consistent approach.}.
This can now be done via two kinds of RPA's;
one with sigma-like excitations and an other   
with pion-like excitations.
For symmetry considerations, we rather favor the second. 
Indeed it was shown in \cite{aoui96} that a RPA with the quantum numbers of the 
pion field can be made symmetry conserving by enlarging the 
RPA excitation operator in such a way to accommodate the whole set of
excitations formally present in the two-body axial charge operator. 
On the other hand, Noether theorem defines the axial current to be
 
\ba
A_{5\,\mu}^a = i\bar{\psi}\gamma_{\mu}\gamma_5 \frac{\tau^a}{2}\psi
 \,+\, \partial_{\mu}\sigma \pi^a -   \partial_{\mu}\pi^a (\sigma + 
 \ave{\hat{\sigma}})~. 
\label{eq20}
\ea

The symmetry operator $Q_5^a$, being the volume integral of the time component
$A_{5\,0}^a$, takes in second quantization a structure 
in which one can identify linear forms standing for the creation and 
annihilation of quasi-pion modes at rest, 
as well as bilinear forms corresponding to creation and annihilation 
of pairs of either fermions or bosons with opposite parities and vanishing 
total momentum. 

According to the Goldstone theorem, the action 
of the axial charge on the full correlated vacuum with a spontaneously
broken phase creates in the chiral limit a non normalizable
state which is nothing but the pion state
\bd
{Q}_5^a|vac\rangle \propto |{\pi}^a\rangle ~.
\ed
Here one can mimic this situation
and build, on an approximate vacuum  
which is considered here to be a RPA ground state $( \ket{RPA})$, a normalizable 
pion state, away from the chiral limit, using the 
following RPA excitation operator

\ba
{Q}_{\nu}^{a\,+} &=& X_{\nu}^{(1)} {\alpha}_{0}^{a\,+} - Y_{\nu}^{(1)} {\alpha}_{0}^{a} 
\nonumber\\
&+& \sum_k \left[ X_{\nu}^{(2)}(k) {\alpha}^{a\,+}_{-k} \beta^+_k  - 
Y_{\nu}^{(2)}(k)  {\alpha}_{-k}^{a} \beta_k  \right] 
\,+ \, 
\sum_k \left[ X_{\nu}^{(3)}(k)  {\alpha}^{a\,+}_{-k} \beta_k  - 
Y_{\nu}^{(3)}(k) {\alpha}_{-k}^{a} \beta_k^+  \right] 
\nonumber \\
&+& 
\sum_{krr'} \left[ X_{\nu}^{(4)}(k) C^+_{kr} D^+_{-kr'} - Y_{\nu}^{(4)}(k) D_{-kr'} C_{kr} \right]
\,+\,
\sum_{krr'} \left[ X_{\nu}^{(5)}(k) C^+_{kr} C_{kr'} - Y_{\nu}^{(5)}(k) D_{kr'} D^+_{kr} \right]
~.
\label{eq22}
\ea
The latter is defined to contain the same pair-excitations as the axial charge 
but with amplitudes which remains to be fixed dynamically. 
The RPA ground state is defined as the vacuum for the above 
operator ${Q}_{\pi}^a\ket{RPA} = 0$, while the single pion state, given by 
${Q}_{\pi}^{a\,+}\ket{RPA} = \ket{\pi^a}$, is normalized as follows

\be
\langle RPA | \left[ Q_{\mu}^{a}, Q^{a\,+}_{\nu} 
\right] | RPA \rangle = \delta_{\mu \nu}~.
\label{eq23}
\ee

Using Rowe's equation of motion method \cite{rowe68}, 
one is lead to the usual RPA secular 
equations
 
\be
\bra{RPA} \left[ \delta Q_{\nu}^{a}, \left[ H, {Q}^{a\,+}_{\nu} \right] 
\right] \ket{RPA}
= \omega_{\nu} \bra{RPA} \left[ \delta {Q}_{\nu}^{a}, 
{Q}^{a\,+}_{\nu} \right] \ket{RPA}~.
\label{eq24}
\ee
These are solved within the usual quasi-boson approximation in which  
the $\ket{RPA}$
ground state is replaced by
the HFB ground state $\ket{\Phi}$. This procedure which seems brutal and inconsistent
is nevertheless commonly used and fully under control. It can be shown
that it is equivalent to a lowest order truncated 
bosonization of all the bilinear operator products which form the Hamiltonian
and the excitation operator.  
In the ansatz above, only the number non-conserving bilinear terms contribute
effectively to the RPA equation Eq.(\ref{eq24}).     
The remaining number conserving terms decouples totally. They should appear in
higher non harmonic orders in the bosonization. This point will not be
developed further here.
The equations of motion Eq.(\ref{eq24}) are in fact a set of coupled channel
equations where all possible states with opposite parities scatter
according to the vertices allowed by the Lagrangian Eq.(\ref{eq1}).
These coupled-channel Lippmann-Schwinger equations are further coupled to
a Dyson equation for the pion propagator. These processes are represented
schematically below 

\begin{center}
$\left(
\begin{array}{ccc}
\quad {\vec \pi} \rightarrow {\vec \pi} \quad &
\quad {\vec \pi} \rightarrow {\vec \pi} \sigma \quad & 
\quad {\vec \pi} \rightarrow \overline{\psi} \gamma_5 \frac{\vec \tau}{2} \psi \quad\\
\quad {\vec \pi} \sigma \rightarrow {\vec \pi} \quad 
&\quad {\vec \pi} \sigma \rightarrow {\vec \pi} \sigma \quad & \quad  0 \quad \\
\quad \overline{\psi} \gamma_5 \frac{\vec \tau}{2}\psi \rightarrow {\vec \pi} \quad &\quad   0 \quad &  
\quad \overline{\psi} \gamma_5 \frac{\vec \tau}{2}\psi \rightarrow
\overline{\psi} \gamma_5 \frac{\vec \tau}{2}\psi\quad 
\end{array} 
\right) $
\end{center}

Since the linear $\sigma$-model does not provide for a coupling between 
pairs of fermions and bosons the matrix which governs the subspace for pure
fermionic re-scattering is actually diagonal. Including quartic interaction 
terms \`a la Nambu Jona-Lasinio leads to a fermionic RPA type of re-scattering.
This is briefly shown at the end of the paper.

Let us turn now to the proper resolution of the RPA equations. These can be 
recast in the following eigenvalue problem form

\begin{equation}
\int d{\vec q}_2
\left(
\begin{array}{cc}
{\cal A}({\vec q}_1,{\vec q}_2) &
{\cal B}({\vec q}_1,{\vec q}_2) \\ 
{\cal B}({\vec q}_1,{\vec q}_2) &
{\cal A}({\vec q}_1,{\vec q}_2)
\end{array} 
\right) 
\left(\begin{array}{c}
{\cal U}_{\nu}({\vec q}_2) \\ {\cal V}_{\nu}({\vec q}_2)
\end{array} \right) 
\, =\, \omega_{\nu} {\cal N}
\left(\begin{array}{c}
{\cal U}_{\nu}({\vec q}_1) \\ {\cal V}_{\nu}({\vec q}_1)
\end{array}\right)~.
\label{eq25}
\end{equation}
As anticipated above, the amplitudes $X^{(3)}$, $Y^{(3)}$, $X^{(5)}$ and $Y^{(5)}$ in Eq.(\ref{eq22}), 
corresponding to 
the excitations generated by all number conserving pairs of operators,
are found to decouple in the present case.
It will be seen later on
that this situation does not persist at finite temperature 
for the number conserving bosonic pairs. The number conserving fermionic 
pairs, on the other hand, 
start contributing only at finite total three momentum 
and finite baryon density. 
Thus in the eigenvalue RPA equation above,   
${\cal A}$ and ${\cal B}$ are $3\times 3$ matrices.   
The ${\cal A}$ matrix is the sum of a diagonal matrix which represents the
free propagation of re-scattering states, and the ${\cal B}$ matrix such
that  
\begin{equation}
{\cal A}({\vec q}_1,{\vec q}_2) = 
\left(
\begin{array}{ccc}
{\cal E}_{\pi}(0) & 0 & 0 \\
0 & \left[{\cal E}_{\pi}({\vec q}_1) + {\cal E}_{\sigma}({\vec q}_1)\right] &0 \\
0 & 0 &  2 {\cal E}_{F}({\vec q}_1) 
\end{array} 
\right) \, \delta({\vec q}_1-{\vec q}_2)
\quad +\quad  {\cal B}({\vec q}_1,{\vec q}_2)~.
\label{eq26}
\end{equation}
The interaction is fully encoded in the ${\cal B}$ matrix which reads 
\begin{equation}
{\cal B}({\vec q}_1,{\vec q}_2) = 
\frac{1}{\sqrt{2 {\cal E}_{\pi}(0)}}
\left(
\begin{array}{ccc}
 0  & 2 \lambda^2 \ave{\hat{\sigma}} \, {\cal R}({\vec q}_2)  &  \gamma g (2\pi)^{-\frac{3}{2}} \\ 
2 \lambda^2 \ave{\hat{\sigma}} \, {\cal R}({\vec q}_1) &
2 \lambda^2 \, {\cal R}({\vec q}_1) {\cal R}({\vec q}_2)&  0 \\
g (2\pi)^{-\frac{3}{2}}  & 0 & 0 
\end{array} 
\right)~, 
\label{eq27}
\end{equation}
with
\begin{equation}
{\cal R}({\vec q}) = \left[(2\pi)^3 4 
{\cal E}_{\pi}({\vec q}){\cal E}_{\sigma}({\vec q}) \right]^{-1/2}~.
\label{eq28}
\end{equation}
Finally the norm matrix  ${\cal N}$ as well as  the RPA eigenvectors
${\cal U}_{\nu}({\vec q})$ and ${\cal V}_{\nu}({\vec q})$ are given by:
\begin{equation}
{\cal N} = \left(
\begin{array}{cc}
I_d & 0 \\
0 & -I_d
\end{array} \right)~, \quad\quad
{\cal U}_{\nu}({\vec q})=\left(
\begin{array}{c}
X_{\nu}^{(1)} \\
X_{\nu}^{(2)}({\vec q}) \\
X_{\nu}^{(4)}({\vec q}) \\
\end{array} \right)~,
\quad\quad
{\cal V}_{\nu}({\vec q})=\left(
\begin{array}{c}
Y_{\nu}^{(1)} \\
Y_{\nu}^{(2)}({\vec q}) \\
Y_{\nu}^{(4)}({\vec q}) \\
\end{array} \right)~,
\label{eq29}
\end{equation}
where $I_d$ is the $3\times 3$ identity matrix.

The solution of the eigenvalue problem proceeds through straightforward 
calculation. This can be carried analytically here via a so called Feshbach 
projection from the two-particle states subspaces onto the single 
pion-state subspace. We get the following expression for the RPA eigenvalues

\begin{equation}
\omega_{\nu}^2 = {\cal E}_{\pi}^2(0) \,\,+\,\, g^2 
\Sigma_{F\bar{F}}(\omega_{\nu}^2)
\,\, + \,\,\frac{4\lambda^4 \,\ave{\hat{\sigma}}^2 \,
\Sigma_{\pi \sigma}(\omega^2_{\nu})}
{1 \,\,-\,\, 2\lambda^2\, \Sigma_{\pi \sigma}(\omega_{\nu}^2)}~,
\label{eq30}  
\end{equation}
where $\Sigma_{F\bar{F}}$ and $\Sigma_{\pi\sigma}$ are the RPA bubbles given by
\begin{eqnarray}
\Sigma_{F\bar{F}}(\omega_{\nu}^2)\,&=&\, \gamma
\int \frac{d{\vec p}}{(2\pi)^3} \, \,
\frac{ 4 {\cal E}_F(p) }{
\omega_{\nu}^2 \,\,-\,\, 4\, {\cal E}_{F}(p)^2}~,
\nonumber\\
\Sigma_{\pi\sigma}(\omega_{\nu}^2)\,&=&\,
\int \frac{d{\vec p}}{(2\pi)^3} 
\frac{{\cal E}_{\pi}(p)+{\cal E}_{\sigma}(p)}
{2 {\cal E}_{\pi}(p){\cal E}_{\sigma}(p)} 
\frac{1}{\omega_{\nu}^2-({\cal E}_{\pi}(p)+{\cal E}_{\sigma}(p))^2}~.
\label{eq31} 
\end{eqnarray}
Eqs.(\ref{eq30},\ref{eq31}) are diagrammatically represented in Fig. (2).
As it was motivated earlier, one of the RPA solutions has to have the Goldstone
character. This solution is commonly known, in the nuclear problem, 
as the spurious solution. This negative connotation expresses the fact that 
this solution does not correspond to a real excitation of
the system, but rather is an expression of a dynamical breaking of a symmetry 
in the system as, for instance, the space rotation or translation of a nucleus.
In the present case, we hold the spurious solution for a real excitation of
the vacuum and it corresponds to the pionic mode. To retrieve explicitly this
solution,  we first notice the following identities 

\begin{equation}
\Sigma_{\pi\sigma}(0)\,=\, \frac{I_{\pi} - I_{\sigma}}
{{\cal E}^{2}_{\pi}(0)-{\cal E}^{2}_{\sigma}(0)} ~,\quad\quad\quad  
\Sigma_{F\bar{F}}(0) \,=\, \frac{I_F}{{\cal E}_F(0)}~.
\label{eq32}  
\end{equation}
These, together with the gap equations Eq.(\ref{eq15}),
allow  to rewrite the RPA frequencies in Eq.(\ref{eq30}) into the 
following form 
\begin{equation}
\omega_{\nu}^2 = \frac{c}{\ave{\hat{\sigma}}} \,+\, 
 g^2 \left[ \Sigma_{F\bar{F}}(\omega_{\nu}^2) - \Sigma_{F\bar{F}}(0)\right]
\,\, + \,\,
\frac{2\lambda^2 \left[{\cal E}_{\sigma}^{2}(0) - {\cal E}_{\pi}^{2}(0) \right]
 \, \left[\Sigma_{\pi \sigma}(\omega^2_{\nu}) -
 \Sigma_{\pi \sigma}(0) \right] }
{1 \,\,-\,\, 2\lambda^2\, \Sigma_{\pi \sigma}(\omega_{\nu}^2)}
\label{eq33}  
\end{equation} 

It is clear, from the expression above, that the Goldstone solution is
manifestly  present in the RPA
spectrum. Besides, there exist 
evidently a continuum of solutions which correspond to two cuts; 
one is the free
propagation of pairs of fermions with opposite parities, and the other
corresponds to the quasi-sigma and quasi-pion scattering process.
 
It is well known from the nuclear many-body problem that the random phase
approximation is the perturbative procedure which  further
diagonalizes the residual interaction inherited from the mean field calculation.  
In the present case, the considered RPA did not affect the whole residual
interaction. Actually, the part of the Hamiltonian which
was diagonalized here is the one which allows the transition between states  
of mixed parities. This part is given by 

\ba
{\cal H}_{RPA} &=& 
\int  d{\vec p}\, {\cal E}_F(p) \sum_{r} 
\left[ C_{{\bf p}r}^+C_{{\bf p}r} +D_{{\bf p}s}^+D_{{\bf p}s} \right] 
+ \int d{\vec p}\, {\cal E}_{\pi}(p) \vec{\alpha}_{p}^+ \vec{\alpha}_{p}
+ \int d{\vec p} \,{\cal E}_{\sigma}(p)  \beta^+_p \beta_p
      \nonumber\\
&+&   \int d{\bf x}  :\left[ \lambda^2 \ave{\hat{\sigma}}\, \sigma {\vec{\pi}}^2  +
 \frac{\lambda^2}{2} {\sigma^2} {\vec{\pi}}^2 
 - ig \bar{\psi}   
{\vec \tau} {\vec \pi} \gamma_5  \psi \right] : ~, 
\label{eq34}
\ea
and the full Hamiltonian reads
\ba
{\cal H} &=& {\cal H}_0 + {\cal H}_{RPA} + {\cal H}_{residual} \nonumber\\
{\cal H}_{residual} &=& 
\int d{\bf x} : \left[ \lambda^2 \ave{\hat{\sigma}} {\sigma^3}  +
 \frac{\lambda^2}{4}\left( (\vec{\pi}^2)^2 + {\sigma^4} \right)
 - g \bar{\psi} \sigma  \psi \right] :  
\label{eq35}
\ea
where ${\cal H}_0$ is the mean field contribution to the ground state energy
and the semicolons stand for the normal ordering with respect to the HFB
squeezed state.
The RPA diagonalization brings as well a finite contribution to the vacuum 
energy. In the RPA basis, the ${\cal H}_{RPA}$ piece of the Hamiltonian has,  
therefore, the following form 
\ba
{\cal H}_{RPA} &=& \, \, E_{RPA}\,\, +\,\, :{\cal H}_{RPA}: \nonumber\\
:{\cal H}_{RPA}:&=& \,\,\sum_{q,\nu} \omega_{\nu}(q) Q_{\nu}^+(q)Q_{\nu}(q)
\nonumber\\
E_{RPA} &=& \,\,\bra{RPA}{\cal H}_{RPA} \ket{RPA} 
\label{eq36}
\ea
where the semicolons indicate this time a normal ordering 
with respect to the RPA ground state. The evaluation of the RPA contribution
to the ground state requires in fact a finite three momentum RPA calculation  
which was not addressed in this paper. Therefore the $E_{RPA}$ will not be
explicitly given
here. This point as well as the nuclear matter equation of state will be
addressed soon in a forthcoming publication. 
As a prerequisite, let us next see how to extend the
present approach to the case of finite temperature and finite baryon density. 
\section{ FINITE TEMPERATURE AND BARYON DENSITY EXTENSIONS}

The extension of the above approach to a baryon rich system at finite
temperature follows through very standard technics. 
One of these is the Thermo Field Dynamics (TFD) \cite{umez82}. 
This approach embraces very well the concept of dynamical mappings 
since it keeps transparent the notion of the vacuum and thus of the Fock space. 
However, it remains very much involved in the present type of calculations 
because it proceeds via a doubling of the dynamical variables ({\it i.e.} 
doubling of the Fock space).
In what follows we adopt the more practical and rather standard approach which 
consists in evaluating the Grand Canonical Potential $\Omega$
of the grand canonical ensemble. The conditions of stationarity (vanishing of 
the first derivative of $\Omega$), and stability (positivity of its second 
derivative) lead to the thermodynamical equilibrium of the system.
In the present case, $\Omega$ takes the form  
\be 
\Omega = \ave{H} \,-\, T S \,-\,   \mu \ave{N}~.  
\label{eq37}
\ee
Given the grand canonical partition function $Z =
Tr[e^{-\beta(H- \mu N)}]$, and density operator 
$D = Z^{-1}[e^{-\beta(H-  \mu N)}]$~, with $\beta = 1/k_B T$, 
the expectation values in Eq.(\ref{eq37}) are short hand notations which 
express the following traces
\be
\ave{H} = Tr[DH]~, \quad\quad  S = \ave{-k_B ln(D)}= Tr[-k_B D ln(D)]~, \quad\quad
\ave{N} = Tr[DN]~.
\label{eq38}
\ee
Here $k_B$ is the Boltzmann constant, $T$ the temperature, ${\mu}$ the baryon chemical
potential,  $H$  the model Hamiltonian 
of the system, and $N$ the baryonic particles
number operator, given as usual by:
\be
N  = \int d{\bf x}\,\, :\bar{\psi}\gamma^0 \psi :~. 
\label{eq39}
\ee   
For a theory of free fields, {\it i.e.} when $H$ reduces to a diagonal
bilinear form in the field operators, computing the thermodynamics of the present
system is a common textbook exercise \cite{fett71}. The partition function $Z$ 
takes then a simple form which allows to express, for instance, the entropy $S$
as \cite{toda85} 
\ba
S =  &-& k_B \sum_{\vec{p},\rho = \pi , \sigma}
\left[ f_{\rho}(\vec{p})lnf_{\rho}(\vec{p}) - 
(1+f_{\rho}(\vec{p}))ln(1+f_{\rho}(\vec{p})) \right] 
\nonumber \\
&-&  k_B \sum_{\vec{p},\rho = F , \bar{F}}
\left[f_{\rho}(\vec{p})lnf_{\rho}(\vec{p}) - 
(1-f_{\rho}(\vec{p}))ln(1-f_{\rho}(\vec{p}))\right]~,  
\label{eq40}
\ea
where $f_{\pi,\sigma}$ are Bose occupation numbers for the pion and the sigma 
modes, respectively, while $f_{F,\bar{F}}$ are Fermi occupation numbers for the 
baryons. For an interacting 
system of particles, the solution for the partition function $Z$ as well as for 
the
entropy $S$, as sketched above, persists in its simple and sympathetic form only
in the case of an independent quasi-particle picture. 
The Hamiltonian, in this case,   is 
split  into a diagonal bilinear part which is fully admitted into 
the thermodynamics, and a residual interaction which is subsequently treated in 
a perturbation.    
There exist several ways of realizing a quasi-particle picture. These are based, 
for instance, on the self-consistent approximations \`a la Hartree-Bogoliubov 
or \`a la Hartree-Fock-Bogoliubov\footnote{Indeed both of them  deliver 
self-consistent operator bases  which can keep the bilinear part 
of the Hamiltonian
diagonal, thus allowing  an independent quasi-particle picture.}. 
Since we have established in the previous sections that, for symmetry 
requirements, the HFB basis is needed, we then chose to realize 
the independent quasi-particle picture in this latter. 
For this purpose, a thermal Bogoliubov 
rotation is applied to each field  in the Hamiltonian according to: 
\ba
\vec{\alpha}_{\bf p}^+(T) &=& u_{\pi}(p,T)\vec{a}_{\bf p}^+ - v_{\pi}(p,T) 
\vec{a}_{\bf
-p}~~;~~~
\beta_{{\bf p}}^+(T) = u_{\sigma}(p,T)b_{{\bf p}}^+ - v_{\sigma}(p,T) b_{{\bf -p}} - 
w_0(T) \delta({\bf p})
\nonumber \\
C_{{\bf p}r}^+(T) &=& u_F(p,T) c_{{\bf p}r}^+ - r v_F(p,T) d_{{\bf -p}r}~~;~~~
D_{{\bf p}r}^+(T) = u_F(p,T) d_{{\bf p}r}^+ + r v_F(p,T) c_{{\bf -p}r}
\label{eq41}
\ea
Here too the canonical normalization of the thermal amplitudes 
$u_{\pi}(p,T)$, $u_{\sigma}(p,T)$, $v_{\pi}(p,T)$, $v_{\sigma}(p,T)$, $u_F(p,T)$
and $v_F(p,T)$ is assumed
\be
u_{\pi}(p,T)^2 - v_{\pi}(p,T)^2 = 1~,\quad
u_{\sigma}(p,T)^2 - v_{\sigma}(p,T)^2 = 1~,\quad
u_F(p,T)^2 + v_F(p,T)^2 = 1.
\label{eq42}
\ee
Except for the temperature dependence, the above transformations strictly follow 
the zero temperature ones performed  previously. 
Using Bloch - De Dominicis theorem \cite{bloc56},
it is straightforward to compute the traces  $\ave{H}$ and $\ave{N}$. 
The entropy
$S$ keeps its form in Eq.(\ref{eq40}) with the occupation factors given now 
in terms of the self-consistent thermal quasi-energies by:
\ba
f_{\pi}(q) &=& \left[ {\mathrm exp} \left(\beta{\cal E}_{\pi}^T(q) \right)\, - 
\,1 \right]^{-1}~,
\quad\quad\quad
 f_{\sigma}(q) = \left[ {\mathrm exp} \left(\beta{\cal E}_{\sigma}^T(q)
 \right)\, - \,1 \right]^{-1}~,
 \nonumber\\
f_F(q) &=& \left[ {\mathrm exp} \left(\beta 
\left({\cal E}_{F}^T(q) -\mu \right) \right) \,+ \,1 \right]^{-1}~,
\quad\quad
f_{\bar{F}}(q) = \left[ {\mathrm exp} \left(\beta
\left({\cal E}_{F}^T(q) +\mu \right) \right) \, +\, 1 \right]^{-1}~.
\label{eq43}
\ea  
The thermal average of H  reads:
\begin{eqnarray}
\ave{H}
  &=& 
\int\frac{d{\vec q}}{(2\pi)^3}\, \frac{\omega_{q}}{2} \left[ 
3\left(1+2f_{\pi}(q)\right)\left(u_{\pi}(q,T)^2 + v_{\pi}(q,T)^2\right) + 
\left(1+2f_{\sigma}(q)\right)\left(u_{\sigma}(q,T)^2 + v_{\sigma}(q,T)^2\right)
\right]
\nonumber\\
&+& \gamma \int \frac{d{\vec q}}{(2\pi)^3} \left( f_F(q)+f_{\bar{F}}(q)-1 \right)
\left[ E_q \left( u_F(q,T)^2 - v_F(q,T)^2 \right) - 
2g \ave{\hat{\sigma}}_T u_F(q,T)v_F(q,T) \right]
\nonumber\\
 &+& \frac{3 \lambda^2}{4} \left[ I_{\sigma}^TI_{\sigma}^T +
 5 I_{\pi}^T I_{\pi}^T + 2 I_{\pi}^T I_{\sigma}^T 
 + 2 \ave{\hat{\sigma}}_T^2\left( I_{\pi}^T + I_{\sigma}^T \right) \right]
- c \ave{\hat{\sigma}}_T  +
\frac{\mu^2 \ave{\hat{\sigma}}_T^2}{2} + \frac{\lambda^2 \ave{\hat{\sigma}}_T^4}{4} 
~,
\label{eq44}
\end{eqnarray}
With the definitions
\begin{equation}
I_{\pi}^T = \int d{\bf x} \,\, \ave{ \pi_i({\bf x}) \pi_i({\bf x}) }~,\quad
I_{\sigma}^T = \int d{\bf x} \,\, \ave{ \sigma({\bf x}) \sigma({\bf x}) }~,\quad
I_{F}^T = \int d{\bf x} \,\, \ave{ \bar\psi({\bf x}) \psi({\bf x}) }~.
\label{eq45}
\end{equation}
Minimizing the grand potential $\Omega$ with respect to the shift parameter
$w_0(T)$ gives an equation for the chiral condensate
\begin{equation}
\frac{\delta \Omega}{\delta \ave{\hat{\sigma}}_T} \,=\,
\frac{u_{\sigma}(0,T) + v_{\sigma}(0,T)}{\sqrt{\mu}}
\left[ 3\lambda^2 \ave{\hat{\sigma}}_T ( I_{\pi}^T +  I_{\sigma}^T ) - g I_F^T +  
\lambda^2 \ave{\hat{\sigma}}_T^3 + \mu^2 \ave{\hat{\sigma}}_T -  c \right]\,=\,0
\label{eq46}
\end{equation}
On the other hand, the variations of $\Omega$ with respect to 
the thermal Bogoliubov amplitudes
$u_{\pi}(q,T)$, $u_{\sigma}(q,T)$ and
$u_F(q,T)$, while 
keeping the unitarity constraints Eq.(\ref{eq42}) satisfied, lead to 

\begin{eqnarray} 
\left.\frac{\delta \Omega}{\delta u_{\pi}(q,T)}\right|_{u_{\pi}^2 -v_{\pi}^2 =1} 
&=& c_{\pi}^T(q) \nonumber\\
&=&\omega_{q} u_{\pi}(q,T)\,v_{\pi}(q,T) +
   \frac{\lambda^2}{2}
   \frac{ \left( u_{\pi}(q,T)+v_{\pi}(T,q) \right)^2}{2\omega_{q}}
    \left[ 5I_{\pi}^T + I_{\sigma}^T +  \ave{\hat{\sigma}}_T^2 \right]\,\,=\,\,0 
\nonumber\\
\left.\frac{\delta \Omega}{\delta u_{\sigma}(q,T)}\right|_{u_{\sigma}^2 -v_{\sigma}^2 =1}
&=& c_{\sigma}^T(q) \nonumber\\
 &=& \omega_{q} u_{\sigma}(q,T)\,v_{\sigma}(q,T) +
 \frac{3 \lambda^2}{2}
   \frac{ \left( u_{\sigma}(q,T)+v_{\sigma}(q,T) \right)^2}{2\omega_{q}}
    \left[ I_{\pi}^T + I_{\sigma}^T +  \ave{\hat{\sigma}}_T^2 \right]\,\,=\,\,0
\nonumber\\ 
\left.\frac{\delta \Omega}{\delta u_F(q,T)}\right|_{u_F^2 + v_F^2 =1}
&=& c_{F}^T(q) \nonumber\\
 &=&    2 E_p u_F(q,T)\,v_F(q,T) +
     g \ave{\hat{\sigma}}_T \left[u_F(q,T)^2 - v_F(q,T)^2\right]
\label{eq47}
\end{eqnarray}
It is interesting to notice that the stationarity conditions listed above 
ensure at the same time the independent particle picture for the finite 
temperature HFB solution, giving by this a consistency to the whole approach
\begin{eqnarray}
\ave{[ \alpha_i(q),[H, \alpha_j(p)]]} &\propto& c^T_{\pi}(q) \delta_{q,p}\delta_{ij} = 0\nonumber\\
\ave{[ \beta(q),[H, \beta(p)]]} &\propto& c^T_{\sigma}(q) \delta_{q,p} = 0\nonumber\\
\ave{\{ D_r(q),\{H, C_s(p)\}\}} &\propto& c^T_F(q) \delta_{q,p}\delta_{rs} = 0  
\label{eq48}
\end{eqnarray}
The thermal quasi-particle energies which correspond to the harmonic 
excitations of the system can be computed from 
\begin{eqnarray}
\ave{[ \alpha_i(q),[H, \alpha_j^+(p)]]} &=& {\cal E}^T_{\pi}(q) \delta_{q,p}\delta_{ij}\nonumber\\
& = & 
\omega_{q} \left[u_{\pi}(q,T)^2 + v_{\pi}(q,T)^2\right] +
   \lambda^2
   \frac{(u_{\pi}(q,T)+v_{\pi}(q,T))^{2}}{2\omega_{q}}
    \left[ 5I_{\pi}^T + I_{\sigma}^T +  \ave{\hat{\sigma}}_T^2 \right]
 \nonumber\\
\ave{[ \beta(q),[H, \beta^+(p)]]} &=& {\cal E}^T_{\sigma}(q) \delta_{q,p}\nonumber\\
&=&  \omega_{q} \left[u_{\sigma}(q,T)^2 + v_{\sigma}(q,T)^2\right] +
 3 \lambda^2
   \frac{(u_{\sigma}(q,T)+v_{\sigma}(q,T))^{2}}{2\omega_{q}}
    \left[ I_{\pi}^T + I_{\sigma}^T +  \ave{\hat{\sigma}}_T^2 \right]
\nonumber\\    
\ave{\{ C_r(q),\{H, C_s^+(p)\}\}} &=& {\cal E}^T_F(q) \delta_{q,p}\delta_{rs}\nonumber\\
&=& E_q \left( u_F(q,T)^2 - v_F(q,T)^2 \right) - 
 2g\ave{\hat{\sigma}}_T u_F(q,T)v_F(q,T)
\label{eq49}
\end{eqnarray}
Comparing the above results with those obtained earlier for the zero temperature case,
one can see that the equations describing both stationarity conditions, namely 
the 
ground state energy for the first and the thermal equilibrium for the second,
have kept the same form, apart of course from the presence of the thermal 
occupation factors in the latter.
One can also check that at the minimum of $\Omega$, 
the quasi-particle energies read
\begin{equation}
{\cal E}^T_{\pi\,,\, \sigma}(q)=  
\omega_q \left[ u_{\pi \,,\, \sigma}(q,T)- v_{\pi\,,\, \sigma}(q,T)\right]^2~,
\quad\quad 
{\cal E}^T_{F}(q)= \sqrt{q ^2 + g^2\,\ave{\hat{\sigma}}_T^2}~,
\end{equation}
which lead as well to the familiar expressions of the thermal tadpoles 
\begin{equation}
I_{\pi\,,\, \sigma}^T= \int\frac{d^3q}{(2\pi)^3} 
    \frac{1 + 2 f_{\pi\,,\, \sigma}(q)}{2\,{\cal E}_{\pi\,,\, \sigma}^T(q)}\, ,
    \quad\quad
I_{F}^T= 2 \gamma {\cal E}_F(0)\int\frac{d^3q}{(2\pi)^3} 
    \frac{f_F(q) + f_{\bar{F}}(q) - 1}{{2 \cal E}_F^T(q)}~.    
\end{equation} 
Finally the finite temperature BCS solutions take the form of the following 
four coupled and self-consistent equations:
\ba
{\cal E}_{\pi}^{T\,2}(0) &=&\mu^2 
+ \lambda^2 \left[ 5I^T_{\pi} +I_{\sigma}^T + \ave{\hat{\sigma}}_T^2 \right]~,
\nonumber \\
{\cal E}_{\sigma}^{T\,2}(0) &=& \mu^2 
+ 3\lambda^2 \left[ I^T_{\pi} +I^T_{\sigma} +\ave{\hat{\sigma}}_T^2 \right]~,
\nonumber \\
\ave{\hat{\sigma}}_T &=& - \frac{c}{-\mu^2} - \frac{g}{- \mu^2} I_F^T   
  + \frac{3\lambda^2 \ave{\hat{\sigma}}_T}{-\mu^2} I_{\pi}^T 
  + \frac{3\lambda^2 \ave{\hat{\sigma}}_T}{-\mu^2} I_{\sigma}^T 
  + \frac{\lambda^2 \ave{\hat{\sigma}}_T^3}{-\mu^2} ~'
\nonumber \\
{\cal E}_F^T(0) &=& -g\ave{\hat{\sigma}}_T~.
\label{eq50}
\ea

As in the zero temperature case, the finite temperature BCS equations
lead to a dynamical mass generation and thus to a violation  of the Goldstone 
theorem. Therefore, one needs to go one step further and consider those
necessary thermal fluctuations present in the residual interaction.
One way of bringing about these effects consists in taking 
the limit of a weakly interacting system which 
leads to a linearization of the thermal TDHF  equations. Such a solution
is known to correspond to the thermal RPA approximation (see ref. \cite{somm83} 
for details). 
The TRPA equation are very similar in form to the zero temperature 
Rowe equations of motion Eq.(\ref{eq24}). The expectation values are, however, 
not realized anymore on the RPA ground state. They correspond, instead, to 
traces taken on the grand canonical ensemble, such that    
\be
\ave{ \left[ \delta \vec{Q}_{\nu}, \left[ H, \vec{Q}^+_{\nu} \right] 
\right]}
= \omega_{\nu}^T \ave{ \left[ \delta \vec{Q}_{\nu}, 
\vec{Q}^+_{\nu} \right]}~.
\label{eq51}
\ee
The evaluation of such traces is a task which again can be handled very well 
by means of the Bloch-De Dominicis theorem.  
The net result is the following eigenvalue problem
\begin{equation}
\int d{\vec q}_2 \,{\cal N}^T({\vec q}_1)
\left(
\begin{array}{cc}
{\cal A}^T({\vec q}_1,{\vec q}_2) &
{\cal B}^T({\vec q}_1,{\vec q}_2) \\ 
{\cal B}^T({\vec q}_1,{\vec q}_2) &
{\cal A}^T({\vec q}_1,{\vec q}_2)
\end{array} 
\right) \, {\cal N}^T({\vec q}_2)
\left(\begin{array}{c}
{\cal U}_{\nu}({\vec q}_2,T) \\ {\cal V}_{\nu}({\vec q}_2,T)
\end{array} \right) 
 = \omega_{\nu}(T) {\cal N}^T({\vec q}_1)
\left(\begin{array}{c}
{\cal U}_{\nu}({\vec q}_1,T) \\ {\cal V}_{\nu}({\vec q}_1,T)
\end{array}\right)~,
\label{eq52}
\end{equation}
where ${\cal A}^T$ and ${\cal B}^T$ are this time $4\times 4$ matrices 
which have only an implicit dependence on temperature inherited from the temperature 
dependence of the thermal HFB basis. They are given by
\begin{equation}
{\cal A}^T({\vec q}_1,{\vec q}_2) = 
\left(
\begin{array}{cccc}
{\cal E}^T_{\pi}(0) & 0 & 0 & 0\\
0 & \left[{\cal E}^T_{\pi}({\vec q}_1) + {\cal E}^T_{\sigma}({\vec q}_1)\right] 
&0 &0 \\
0 & 0 & \left[{\cal E}^T_{\pi}({\vec q}_1) - {\cal E}^T_{\sigma}({\vec q}_1)\right]
&0 \\
0 & 0 & 0& 2 {\cal E}^T_{F}({\vec q}_1) 
\end{array} 
\right) \, \delta({\vec q}_1-{\vec q}_2)
\, +\,  {\cal B}^T({\vec q}_1,{\vec q}_2) 
\label{eq53}
\end{equation}
The effect of the residual interaction which is responsible for the RPA re-scattering 
is  encoded in the ${\cal B}^T$ matrix which takes the form  
\begin{equation}
{\cal B}^T({\vec q}_1,{\vec q}_2) = 
\frac{1}{\sqrt{2 {\cal E}^T_{\pi}(0)}}
\left(
\begin{array}{cccc}
 0  & 2 \lambda^2 \ave{\hat{\sigma}}_T \, {\cal R}^T({\vec q}_2)  & 
      2 \lambda^2 \ave{\hat{\sigma}}_T \, {\cal R}^T({\vec q}_2)  & 
      \gamma g (2\pi)^{-\frac{3}{2}}\\ 
2 \lambda^2\ave{\hat{\sigma}}_T \, {\cal R}^T({\vec q}_1) &
2 \lambda^2 \, {\cal R}^T({\vec q}_1) {\cal R}^T({\vec q}_2)& 
2 \lambda^2 \, {\cal R}^T({\vec q}_1) {\cal R}^T({\vec q}_2)& 0 \\
2 \lambda^2\ave{\hat{\sigma}}_T \, {\cal R}({\vec q}_1) &
2 \lambda^2 \, {\cal R}^T({\vec q}_1) {\cal R}^T({\vec q}_2)& 
2 \lambda^2 \, {\cal R}^T({\vec q}_1) {\cal R}^T({\vec q}_2)& 0 \\
g (2\pi)^{-\frac{3}{2}} & 0 & 0 & 0
\end{array} 
\right) 
\label{eq54}
\end{equation}
Here ${\cal R}^T({\vec q})$ has an implicit temperature dependence 
and is of the same form as in Eq.(\ref{eq28}).
 
The explicit temperature dependence of the RPA equations is carried by the norm matrix
${\cal N}^T$. This as well as  the RPA amplitudes
${\cal U}_{\nu}({\vec q},T)$ and ${\cal V}_{\nu}({\vec q},T)$ are given by:
\begin{equation}
{\cal N}^T({\vec q}) = \left(
\begin{array}{cc}
N_d^T({\vec q}) & 0 \\
0 & -N_d^T({\vec q})
\end{array} \right)~, \quad\quad
{\cal U}_{\nu}({\vec q},T)=\left(
\begin{array}{c}
X_{\nu}^{(1)}({\vec q},T) \\
X_{\nu}^{(2)}({\vec q},T) \\
X_{\nu}^{(3)}({\vec q},T) \\
X_{\nu}^{(4)}({\vec q},T)
\end{array} \right)~,
\quad\quad
{\cal V}_{\nu}({\vec q},T)=\left(
\begin{array}{c}
Y_{\nu}^{(1)}({\vec q},T) \\
Y_{\nu}^{(2)}({\vec q},T) \\
Y_{\nu}^{(3)}({\vec q},T) \\
Y_{\nu}^{(4)}({\vec q},T)
\end{array} \right)~.
\label{eq55}
\end{equation}
Now $N_d^T$ is a $4 \times 4$ diagonal matrix, having as diagonal elements: 
\begin{eqnarray}
N_d^T({\vec q})_{11} &=& 1 ~,\nonumber\\
N_d^T({\vec q})_{22} &=& 1+ f_{\pi}(q) + f_{\sigma}(q)~, 
\nonumber\\
N_d^T({\vec q})_{33} &=&
 f_{\pi}(q) - f_{\sigma}(q) ~,\nonumber\\
N_d^T({\vec q})_{44} &=&
 1 - f_F(q) - f_{\bar{F}}(q)~.
\label{eq56}
\end{eqnarray}
The difference between $N_d^T$ and $I_d$, the analog matrix at zero temperature
Eq.(\ref{eq29}), represents the single formal departure of the thermal RPA equations from 
the zero temperature ones. Finally one can proceed to the resolution of the RPA 
eigenvalue problem. The RPA frequencies, as in the zero temperature case,
can be read from the characteristic equation 

\begin{equation}
\omega_{\nu}^2(T) = {\cal E}_{\pi}^{T\,2}(0) \,\,+\,\, 
g^2 \Sigma_{F\bar{F}}^T(\omega_{\nu}^2(T))
\,\, + \,\,\frac{4\lambda^4 \,\ave{\hat{\sigma}}_T^2 \,
\Sigma^T_{\pi \sigma}(\omega^2_{\nu}(T))}
{1 \,\,-\,\, 2\lambda^2\, \Sigma_{\pi \sigma}^T(\omega_{\nu}^2(T))}~,
\label{eq57}  
\end{equation}
where 
\begin{eqnarray}
\Sigma_{F\bar{F}}^T(\omega_{\nu}^2(T))\,&=&\, \gamma
\int \frac{d{\vec p}}{(2\pi)^3} \, 4 {\cal E}_F^T(p)\,
\frac{  1 - f_F(p) - f_{\bar{F}}(p)}{
\omega_{\nu}^2(T) \,\,-\,\, 4\, {\cal E}^T_{F}(p)^2}~,
\nonumber\\
\Sigma_{\pi\sigma}^T(\omega_{\nu}^2(T))\,&=&\,
\int \frac{d{\vec p}}{(2\pi)^3} \left[
\frac{{\cal E}^T_{\pi}(p)+{\cal E}^T_{\sigma}(p)}
{2 {\cal E}^T_{\pi}(p){\cal E}^T_{\sigma}(p)} 
\frac{1+ f_{\pi}(p)+f_{\sigma}(p)}{
\omega_{\nu}^2(T)-({\cal E}^T_{\pi}(p)+{\cal E}^T_{\sigma}(p))^2}
\right.
\nonumber\\
& & \quad\quad\quad\quad\quad\quad\quad\quad\,+\,
\left. 
\frac{{\cal E}^T_{\pi}(p)-{\cal E}^T_{\sigma}(p)}{2 {\cal E}^T_{\pi}(p){\cal E}^T_{\sigma}(p)}
\frac{f_{\sigma}(p)-f_{\pi}(p)}{ 
\omega_{\nu}^2(T)-({\cal E}^T_{\pi}(p)-{\cal E}^T_{\sigma}(p))^2 }~.
\right]
\label{eq58} 
\end{eqnarray}
Now, like in the zero temperature case and making use of the identities
\begin{equation}
\Sigma_{\pi\sigma}^T(0)\,=\, \frac{I_{\pi}^T - I_{\sigma}^T}
{{\cal E}^{T\,2}_{\pi}(0)-{\cal E}^{T\,2}_{\sigma}(0)} ~,\quad\quad\quad  
\Sigma_{F\bar{F}}^T(0) \,=\, \frac{I_F^T}{{\cal E}_F^T(0)}~,
\label{eq59}  
\end{equation}
one gets, after inspection of the gap equations, the following
\begin{equation}
\omega_{\nu}^2(T) = \frac{c}{\ave{\hat{\sigma}}_T} + 
 g^2 \left[ \Sigma_{F\bar{F}}^T(\omega_{\nu}^2(T)) - \Sigma_{F\bar{F}}^T(0)\right]
 + 
\frac{2\lambda^2 \left[{\cal E}_{\sigma}^{T\,2}(0) - 
{\cal E}_{\pi}^{T\,2}(0) \right]
  \left[\Sigma^T_{\pi \sigma}(\omega^2_{\nu}(T)) -
 \Sigma^T_{\pi \sigma}(0) \right] }
{1 \,\,-\,\, 2\lambda^2\, \Sigma_{\pi \sigma}^T(\omega_{\nu}^2(T))}~.
\label{eq60}  
\end{equation} 
Equation (\ref{eq60}) exhibits clearly 
a zero energy solution for all temperatures below the transition to the 
Wigner Weyl phase. At the transition and beyond ($\ave{\hat{\sigma}}=0$), 
this solution
is not normalizable, as can be easily seen from the norm matrix ${\cal N}^T$.  

Before closing this paper, it is certainly worth briefly showing how robust
is the field-to-current mapping in systematically preserving 
the symmetry. A  situation of particular interest in the nuclear 
problem is the renormalization of the fermionic four-point function by means 
of the so-called short range correlation. This can be addressed via a chiral
invariant fermionic quartic interaction of Nambu-Jona-Lasinio (NJL) type 
        
\begin{equation}
{\cal L}_1 =  {\cal L}_{\sigma} \,+\, 
 G \left[ \left(\bar{\psi}\psi\right)^2 \,+\,   
\left(\bar{\psi} i \gamma_5 \vec{\tau} \psi \right)^2 \right]~,                       
\label{eq61}
\end{equation} 
where the NJL-coupling G plays  the role of the $g'$ Migdal-parameter. 
The BCS gap equations  are now further coupled by the insertion of a fermionic 
tadpole contribution to the fermion quasi-mass such that 

\ba
{\cal E}_{\pi}^{T\,2}(0) &=&\mu^2 
+ \lambda^2 \left[ 5I^T_{\pi} +I_{\sigma}^T + \ave{\hat{\sigma}}_T^2 \right]~,
\nonumber \\
{\cal E}_{\sigma}^{T\,2}(0) &=& \mu^2 
+ 3\lambda^2 \left[ I^T_{\pi} +I^T_{\sigma} +\ave{\hat{\sigma}}_T^2 \right]~,
\nonumber \\
\ave{\hat{\sigma}}_T &=& - \frac{c}{-\mu^2} - \frac{g}{- \mu^2} I_F^T   
  + \frac{3\lambda^2 \ave{\hat{\sigma}}_T}{-\mu^2} I_{\pi}^T 
  + \frac{3\lambda^2 \ave{\hat{\sigma}}_T}{-\mu^2} I_{\sigma}^T 
  + \frac{\lambda^2 \ave{\hat{\sigma}}_T^3}{-\mu^2} ~,
\nonumber \\
{\cal E}_F^T(0) &=& -g\ave{\hat{\sigma}}_T \,+\, {\tilde G}\, I_F^T~,
\label{eq62}
\ea 

where ${\tilde G} = - 2G(1+ \frac{1}{\gamma})$.
Here again, the asymptotic Goldstone pion is generated  
by the previously introduced parity-mixing RPA. 
This time one gets the following

\begin{eqnarray}
\omega_{\nu}^2(T) = \frac{c}{\ave{\hat{\sigma}}_T} \,&+&\, 
 \frac{g^2 \,
\left[  \Sigma_{F\bar{F}}^T(\omega_{\nu}^2(T)) - \Sigma_{F\bar{F}}^T(0) \right]}
 {\left[ 1 \,\,-\,\,  {\tilde G}\, \Sigma_{F\bar{F}}^T(\omega_{\nu}^2(T))\right]
\left[ 1 \,\,-\,\,  {\tilde G} \,\Sigma_{F\bar{F}}^T(0) \right]}\nonumber\\
& \,+\, &
\frac{2\lambda^2 \left[{\cal E}_{\sigma}^{T\,2}(0) - 
{\cal E}_{\pi}^{T\,2}(0) \right]
 \, \left[\Sigma^T_{\pi \sigma}(\omega^2_{\nu}(T)) -
 \Sigma^T_{\pi \sigma}(0) \right] }
{1 \,\,-\,\, 2\lambda^2\, \Sigma_{\pi \sigma}^T(\omega_{\nu}^2(T))}~.
\label{eq63}  
\end{eqnarray}  
Figures (3) give the diagrammatic representation of the HFB-RPA dynamics in
this case. Of course, further extensions of the model to other 
chirally invariant boson-fermion couplings are possible.

%

%
\section{CONCLUSION}

In conclusion, we have presented an extension of the field-to-current 
mapping, introduced earlier, to a baryon rich regime  at finite temperature. 
The mapping was made dynamical in the Hartree-Fock-Bogoliubov-RPA
approximation. The Goldstone theorem was fulfilled exactly, although
the dynamics was not sorted-out according to neither an expansion in the 
available coupling constants ($\lambda, g, G$) nor in the arbitrary 
charge $N$ and flavor $N_F$ numbers. In fact, it can be
easily appreciated, from the four coupled BCS equations for instance, 
that this approach is highly order mixing. 

All over this work, we did not consider  the p-wave renormalization 
of the pion since it is well known that it does not endanger 
the Goldstone nature of the pion. 
Instead, we fixed our attention on the subtle situation of the  s-wave 
renormalization where potential artefacts, linked to the dynamical mass 
generation, may alter the Goldstone character of the pion. This was indeed
visible from the BCS solution in which all the quasi-particle states were
massive. 
Therefore it is certainly legitimate to state here that the HFB ground-state
is not a viable vacuum and should not be tolerated as an approximate
ground-state for nuclear matter saturation studies for instance. 
This point, although conceptually  crucial and certainly quantitatively
important, is usually not observed in the literature.       
As an alternative, we have presented  the RPA ground-state,  as 
a qualitatively correct approximate ground-state, in which the Nambu-Goldstone
phase is exactly realized. On the other hand, we          
have also indicated that this approach can support whatever 
chirally invariant refinement made to the interaction Lagrangian. 
We hope that this will give  a solid platform
for a program of a  quantitative assessment  of the 
nuclear equation of state, beyond the usually considered
classical mean-field calculations in QHD models.
It should be also mentioned that the present work is
of some relevance to the undergoing quantum mean-field 
calculations, based on chirally-invariant contact nuclear-forces,
undertaken by Nikolaus, Hoch and Madland \cite{niko92}. According to 
these authors, the inclusion of the pion dynamical contributions 
is a step which one needs to envisage seriously in the future \cite{madl97}. 
The non-linear chiral realization \`a la heavy baryon ChPT is certainly
not well suited for a non-perturbative treatment of the dynamics. 
Therefore we see in the present prescription an interesting framework
for such a program.

\acknowledgments
We would like to thank G. Chanfray, W. Greiner, P. Schuck, H. St\"{o}cker and 
J. Wambach for their interest in this work and for their continuous support. We
acknowledge as well the financial support from GSI-Darmstadt.


\vspace*{3.cm}
\begin{figure}
\centerline{\epsfig{file=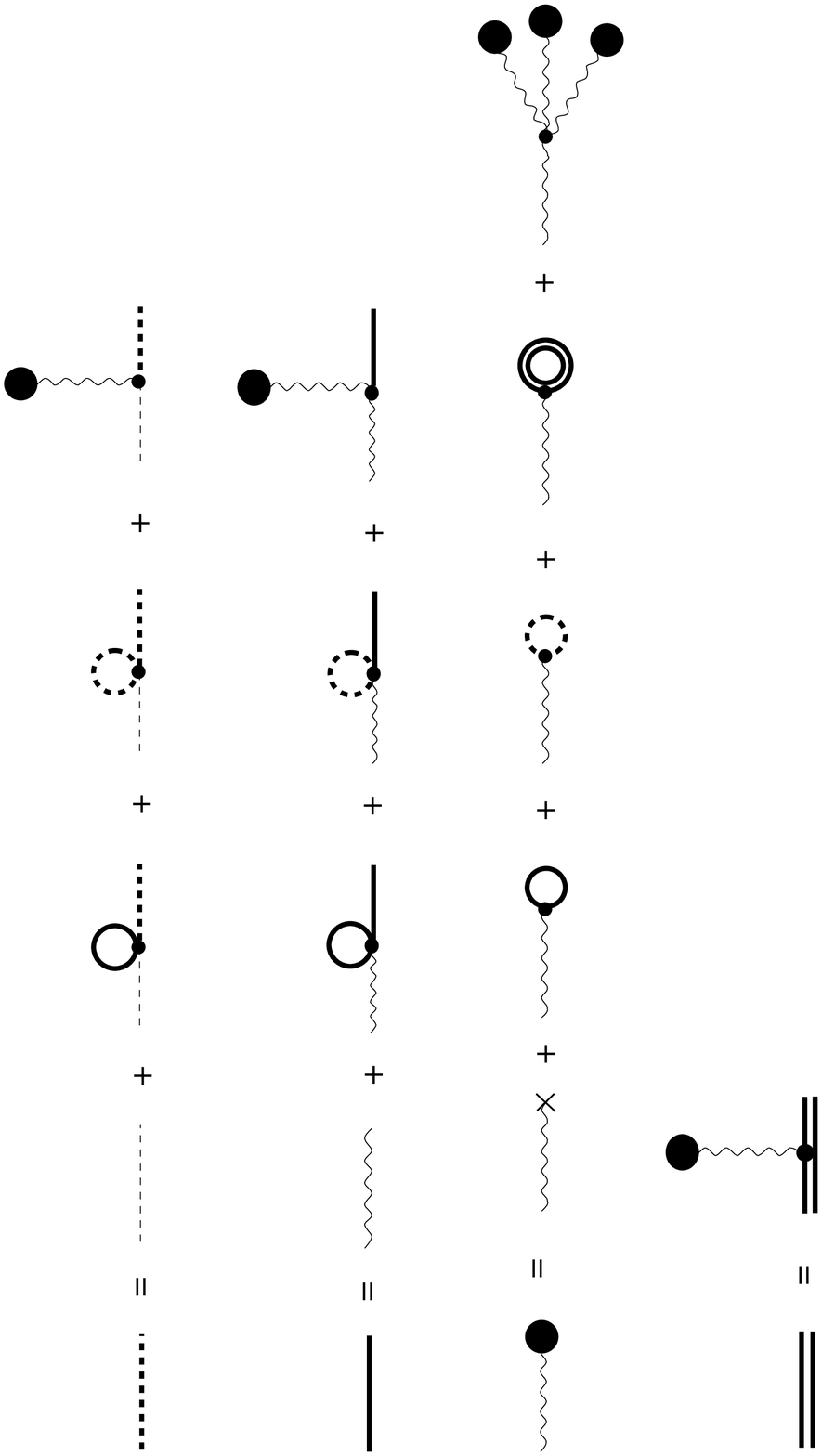,height=16cm,angle=270}}
\vspace*{2.cm}
\caption{Diagrammatic representation of the four coupled BCS solutions. 
The dashed line denotes the self-consistent quasi-pion
propagator, the solid line the quasi-sigma, the double solid line
the quasi-fermion, and  the wavy line the two point 
Green's function of the bare field ${\hat \sigma}$ of 
the Lagrangian density in Eq.(\ref{eq1}).}
\end{figure}

\newpage
\vspace*{3.cm}
\begin{figure}
\centerline{\epsfig{file=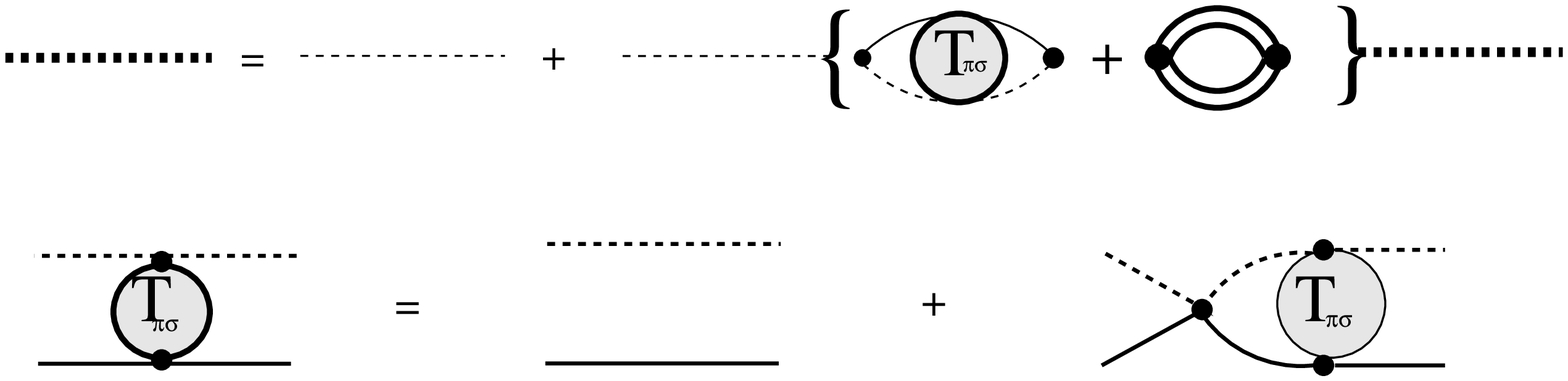,height=4cm,angle=0}}
\vspace*{2.cm}
\caption{Upper part:  The Dyson equation for the physical pion (thick dashed lines) 
for which the mass operator has been extracted from the scattering of the 
quasi-particles in a RPA equation.
Lower part: The scattering equation for a pair of quasi-sigma 
(thin full lines) and quasi-pion (thin dashed lines).}
\end{figure}

\newpage
\vspace*{3.cm}
\begin{figure}
\centerline{\epsfig{file=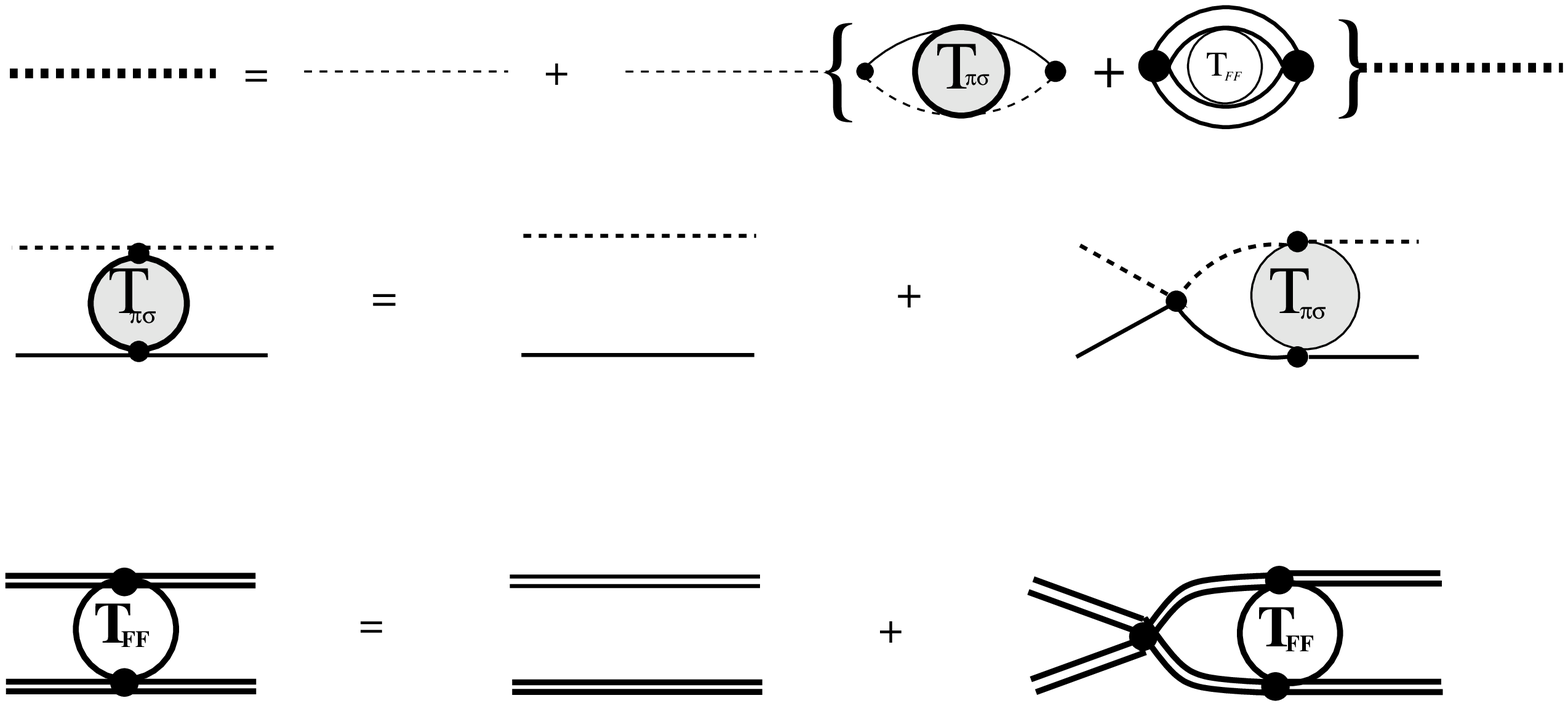,height=6cm,angle=0}}
\vspace*{1.cm}
\centerline{\epsfig{file=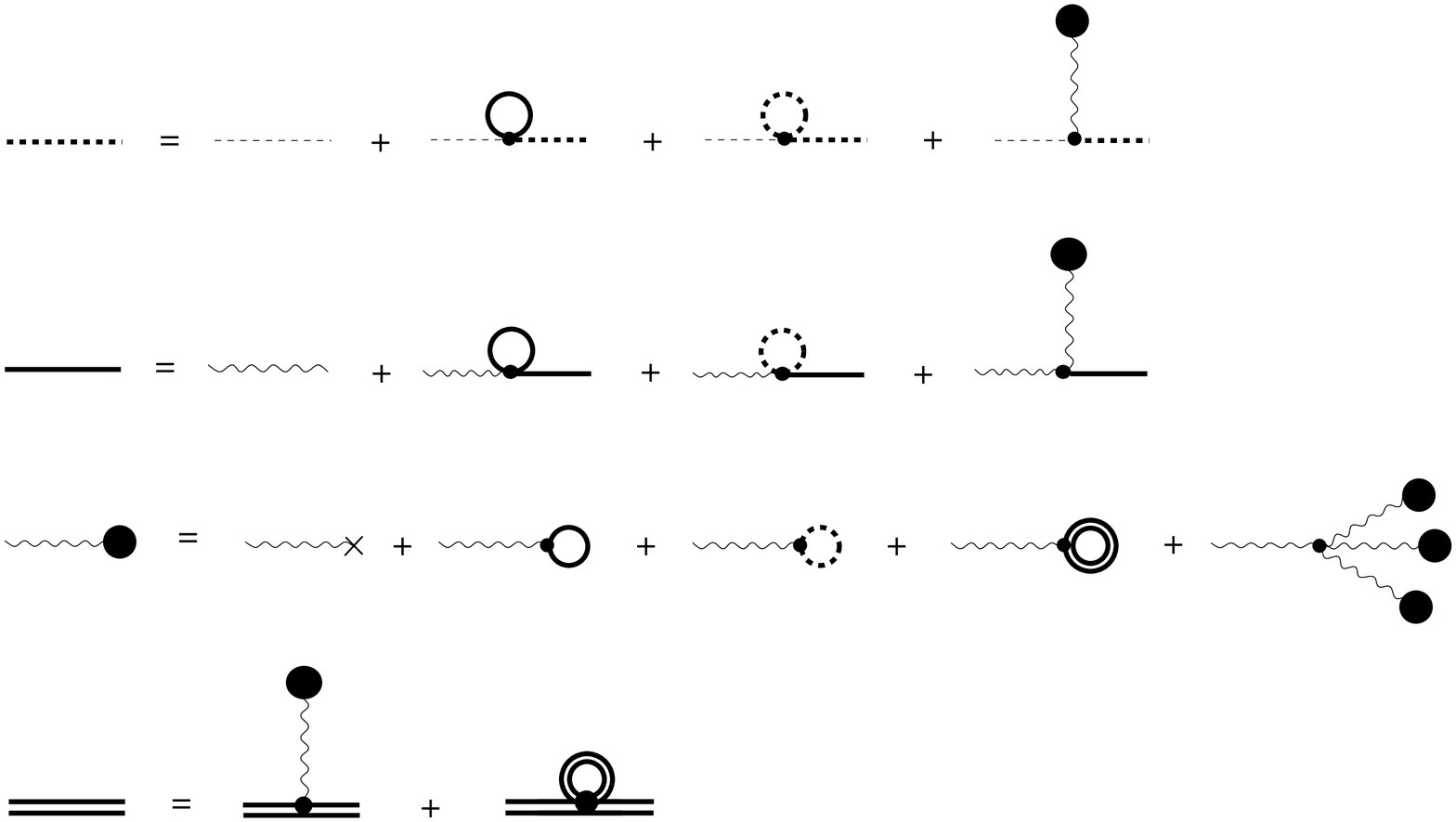,height=8cm,angle=0}}
\vspace*{2.cm}
\caption{The full class of diagrams for the asymptotic Goldstone pion
in the context of the extension discussed in the text.}
\end{figure}

\end{document}